\documentclass[conference]{IEEEtran}
\IEEEoverridecommandlockouts
\usepackage{cite}
\usepackage{amsmath,amssymb,amsfonts}
\usepackage{algorithmic}
\usepackage{graphicx}
\usepackage{textcomp}
\usepackage{xcolor}
\usepackage{multirow}
\usepackage{booktabs}

\def\BibTeX{{\rm B\kern-.05em{\sc i\kern-.025em b}\kern-.08em
    T\kern-.1667em\lower.7ex\hbox{E}\kern-.125emX}}
\begin{document}

\title{Robust HRRP Recognition under Interrupted Sampling Repeater Jamming using a Prior Jamming Information-Guided Network}

\author{\IEEEauthorblockN{1\textsuperscript{st} Guozheng Sun, 2\textsuperscript{nd} Lei Wang*, 3\textsuperscript{rd} Yanhao Wang, 4\textsuperscript{th} Jie Wang, 5\textsuperscript{th} Yimin Liu}
\IEEEauthorblockA{\textit{Department of Electronic Engineering}, \textit{Tsinghua University}\\
Beijing, China \\ sgz25@mails.tsinghua.edu.cn, leiwangqh@tsinghua.edu.cn}}

\maketitle

\begin{abstract}
Radar automatic target recognition (RATR) based on high-resolution range profile (HRRP) has attracted increasing attention due to its ability to capture fine-grained structural features. However, recognizing targets under electronic countermeasures (ECM), especially the mainstream interrupted-sampling repeater jamming (ISRJ), remains a significant challenge, as HRRPs often suffer from serious feature distortion. To address this, we propose a robust HRRP recognition method guided by prior jamming information. Specifically, we introduce a point spread function (PSF) as prior information to model the HRRP distortion induced by ISRJ. Based on this, we design a recognition network that leverages this prior through a prior-guided feature interaction module and a hybrid loss function to enhance the model’s discriminative capability. With the aid of prior information, the model can learn invariant features within distorted HRRP under different jamming parameters. Both the simulated and measured-data experiments demonstrate that our method consistently outperforms state-of-the-art approaches and exhibits stronger generalization capabilities when facing unseen jamming parameters.
\end{abstract}


\section{Introduction}
\label{sec:intro}

Radar automatic target recognition (RATR) is a technique used to identify the category of a target after it has been detected by radars. It typically relies on the target's time-domain or frequency-domain features, such as high-resolution range profile (HRRP)~\cite{8756029}, micro-motion~\cite{11156139}, inverse synthetic aperture radar (ISAR) images~\cite{wang2019Phase} and so on. Among these, HRRP represents the projection of a target’s scattering centers along the radar’s line of sight. Due to its advantages in rapid generation and its ability to capture rich structural details, HRRP has drawn significant attention in the field of RATR. Existing HRRP recognition algorithms can be broadly classified into two categories: traditional recognition methods~\cite{10460189} and deep learning-based recognition methods~\cite{wan2019convolutional}.

\begin{figure}[htbp]
    \centering
    \includegraphics[width=8.3cm,height=8.5cm]{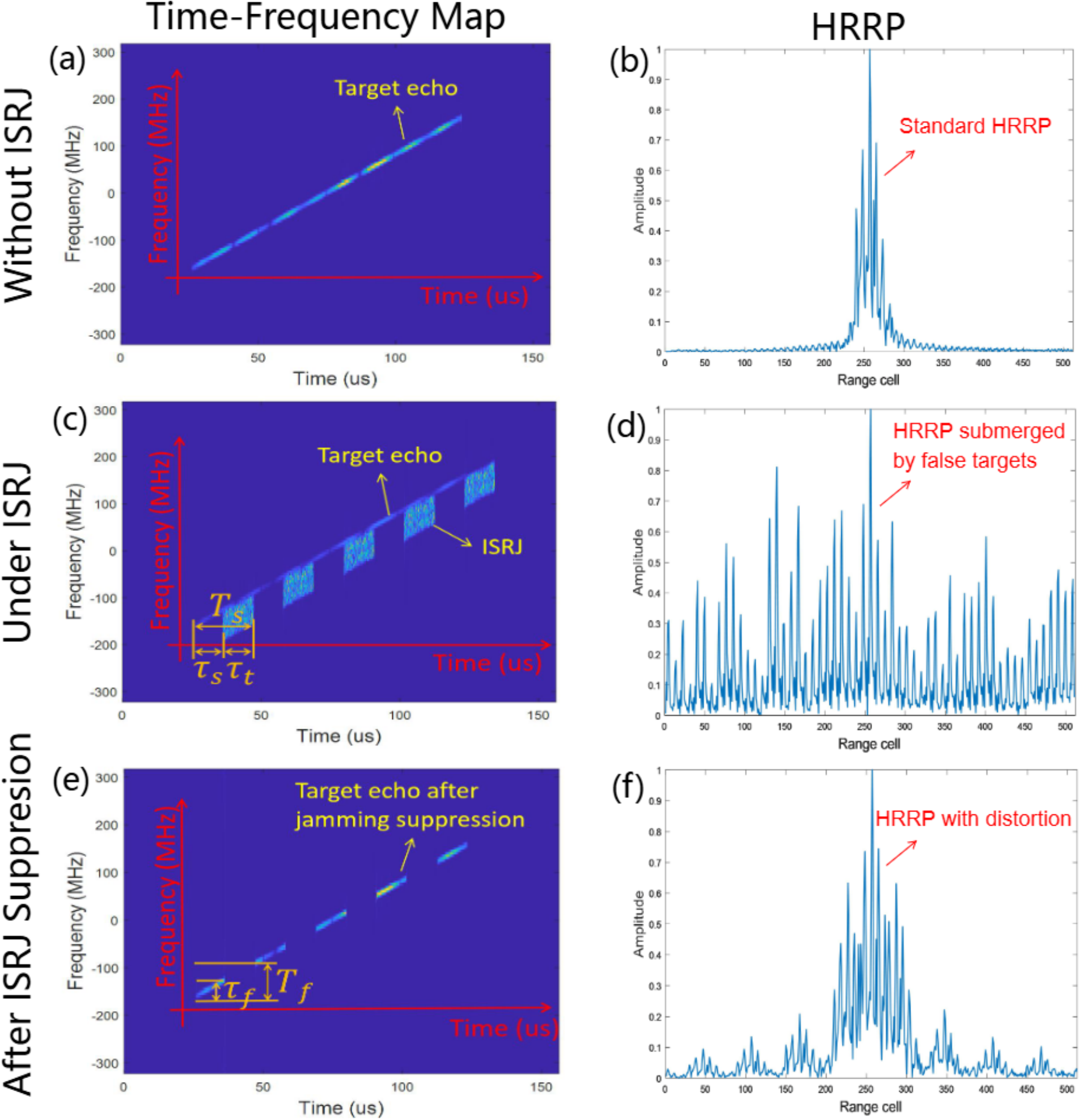}
    \caption{Time–frequency representations and corresponding HRRPs of radar echoes under different conditions.}
    \label{fig:TF}
\end{figure}

Traditional methods generally follow a three-stage pipeline: data preprocessing, feature extraction, and classifier design~\cite{li1993using}. Since the recognition performance largely depends on the accuracy of feature extraction, traditional research has focused on designing effective feature descriptors. However, these methods heavily rely on expert domain knowledge and suffer from high computational complexity, which limits their practicality in real-world scenarios.

With the rapid progress of artificial intelligence, deep learning has emerged as a powerful paradigm for HRRP recognition. Existing methods can be broadly categorized into three groups based on the model architecture: CNN-based models~\cite{wan2019convolutional}, RNN-based models~\cite{9057689}, and Transformer-based models~\cite{gao2024interpretable}, all of which have significantly advanced recognition performance.
Recently, knowledge-guided neural networks have attracted increasing attention in the artificial intelligence community due to their ability to integrate domain-specific priors into the learning process~\cite{meng2025physics}. In the field of HRRP recognition, these networks have also been applied to incorporate physical priors into network architectures, thereby improving recognition performance~\cite{10399838}.
For example, Li \textit{et al.}~\cite{10633835} introduced a prior-assisted gated recurrent unit (GRU) that integrates azimuth, elevation, and range information with HRRP data, thereby improving recognition performance in few-shot scenarios. 

Despite notable progress in HRRP-based target recognition, the presence of electronic countermeasures (ECM) continues to impose significant challenges. Among various ECM techniques, interrupted-sampling repeater jamming (ISRJ) is particularly disruptive due to its ability to inject numerous false targets and introduce severe distortions into the HRRP~\cite{7858673}. As shown in Figure~\ref{fig:TF}d, these distortions destroy the intrinsic range structure and scattering patterns, which are critical for reliable recognition, thereby leading to serious degradation in recognition performance. To the best of our knowledge, there is currently no effective solution for robust target recognition under jamming conditions.

To address this challenge, we propose a novel recognition network that incorporates prior information of jamming characteristics to recognize HRRP under jamming conditions for the first time.
Specifically, in the scenario of ISRJ, we introduce a prior information representation method using the point spread function (PSF) to model the characteristics of jamming. This prior representation effectively captures the distortion effects imposed on HRRP and serves as an indicator of HRRP distortion, providing a solid foundation for robust recognition under ISRJ. Furthermore, the proposed network is equipped with a feature interaction module guided by prior information that integrates HRRP features with the prior jamming features. With the guidance of prior information, the network can learn invariant features within distorted HRRP under various jamming parameters. In addition, a hybrid loss function based on supervised contrastive loss is also employed to enhance the model’s feature discriminability. Collectively, the proposed network exhibits superior robustness and generalization performance when facing unseen jamming parameters.

This paper is organized as follows: Sec.\ref{sec:signal model} presents the signal model of HRRP generating mechanism and the distortion characteristics of HRRP under ISRJ. Sec.\ref{sec:method} introduces the proposed HRRP recognition network under ISRJ. Experiments are presented in Sec.\ref{sec:experiments}, and the conclusion of this paper is drawn in Section \ref{sec:conclusion}.

\begin{figure*}[htbp]
    \centering
    \includegraphics[width=1\linewidth]{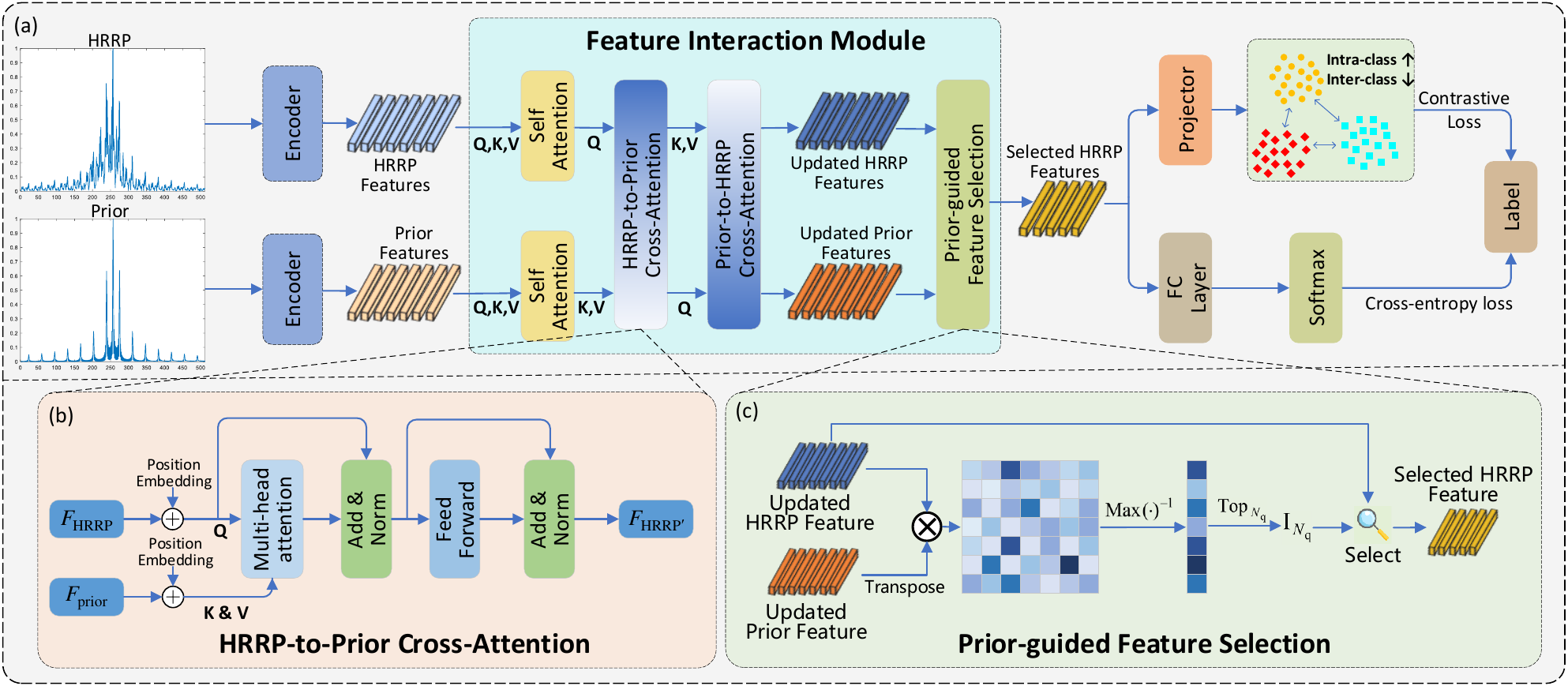}
        \caption{Overall framework of the proposed HRRP recognition network. }
    \label{fig:main}
\end{figure*}

\section{SIGNAL MODEL}
\label{sec:signal model}

\subsection{Formulation of HRRP}

Consider that the radar transmits a linear frequency-modulated (LFM) waveform, as follows:
\begin{equation}
s(t) = a(t) \cdot e^{j2\pi f_0 t}= \text{rect}\left(\frac{t}{T}\right)  e^{j\pi \frac{B}{T} t^2}e^{j2\pi f_0 t} ,
\label{eq:lfm_modulated}
\end{equation}
where \(a(t)\) denotes the baseband signal, \(\text{rect}(t/T)\) is a rectangular window
function with pulse width \(T\), \(B\) is the signal bandwidth, and \(f_0\) is the carrier frequency. Based on Equation~(\ref{eq:lfm_modulated}), the frequency-domain representation of the demodulated echo becomes:
\begin{equation}
S(f) = \sum_k \frac{\sigma_k}{2\pi} A(f) e^{-j2\pi(f + f_0)\frac{2 R_k}{c}},
\label{eq:st_clean}
\end{equation}
where \(A(f)\) denotes the Fourier transform of the LFM waveform, and \(c\) is the speed of light. In addition, \(\sigma_k\) represents the complex scattering coefficient of the \(k\)-th scatterer, and \(R_k\) denotes the distance from the 
\(k\)-th scatterer to the radar. Matched filtering is subsequently performed on Equation~(\ref{eq:st_clean}) by multiplying it with the complex conjugate of \(A(f)\), yielding:
\begin{small}
  \begin{equation}
\begin{split}
    Y(f) = S(f)A^*(f)
         = \sum_{k} A_k \, \text{rect}\left( \frac{f}{B} \right) e^{-j 2\pi (f + f_0)  \frac{2R_k}{c}},
\end{split}
\label{eq:compressed}
\end{equation}  
\end{small}
where \(A_k\) denotes the amplitude of the corresponding scatterer. After applying the inverse Fourier transform, the time-domain expression of the HRRP can be obtained: 

\begin{equation}
   y(t) = 
           \sum_{k} {A}_k B\, \text{Sinc} \left( B \left( t - \frac{ R_k}{c} \right) \right) 
           e^{-j\frac{4\pi f_0 R_k}{c}}.
\label{eq:HRRP}
\end{equation}

\subsection{Formulation of HRRP under ISRJ}

 ISRJ operates by intermittently sampling and retransmitting the radar signal, with a sampling period \(\tau_s\) and a retransmission period \(\tau_t = K \tau_s,\; K \in \mathbb{N}^+\)~\cite{11205831}. In many cases, multiplicative noise is also superimposed onto the signal, further increasing the complexity of the jamming and degrading the features of the echo, as shown in Figure~\ref{fig:TF}c. As can be observed from this figure, it is challenging to achieve effective jamming suppression through methods such as time–frequency analysis. 
 
 In practical applications, jamming suppression based on narrow pulse elimination is typically applied by zeroing out high-energy jamming samples in the echo~\cite{yuan2017method}. As shown in Figure~\ref{fig:TF}d, this process is equivalent to multiplying the frequency-domain echo \(S(f)\) by a jamming suppression operator, which can be modeled as a periodic rectangular window function with frequency-domain period \( T_f \) and frequency-domain pulse width \( \tau_f \). The operator can be expressed as:
\begin{equation}
J(f) = \sum_{n=-\infty}^{+\infty} \operatorname{rect} \left( \frac{f - nT_f}{\tau_f} \right).
\label{eq:J(f)}
\end{equation}

After jamming suppression, the frequency-domain echo becomes:

\begin{equation}
 S'(f) 
=  J(f) \cdot \sum_k \frac{\sigma_k}{2\pi} A(f) e^{-j2\pi(f+f_0) \frac{2R_k}{c}}. 
\label{eq:st_jammed}
\end{equation}

So the matched filtering result of \(S'(f)\) is $Y'(f) = S'(f) \cdot A^*(f) 
    = S(f) \cdot A^*(f) \cdot J(f) = Y(f) \cdot J(f)$.
Then the HRRP under ISRJ is obtained via the inverse Fourier transform of \(Y'(f)\):

\begin{small}
\begin{equation}
\begin{split}
  y'(t) = \mathcal{F}^{-1}\{Y'(f)\} 
  =\sum_{n=-\infty}^{+\infty} A' y(t - \frac{n}{T_f}) \text{Sinc}\left(\frac{n\tau_f}{ T_f}\right),
\end{split}
\label{eq:hrrp_conv}
\end{equation}  
\end{small}
where \(y(t)\) denotes the HRRP in the absence of jamming shown in Equation~(\ref{eq:HRRP}), and \(A'\) represents the amplitude. It can be observed that the HRRP after applying jamming suppression exhibits periodic grating lobes, which degrade the recognition performance.

\section{METHODOLOGY}
\label{sec:method}

\subsection{Prior Jamming Information Modeling Method}

\begin{figure}[htbp]
    \centering
    \includegraphics[width=0.45\textwidth]{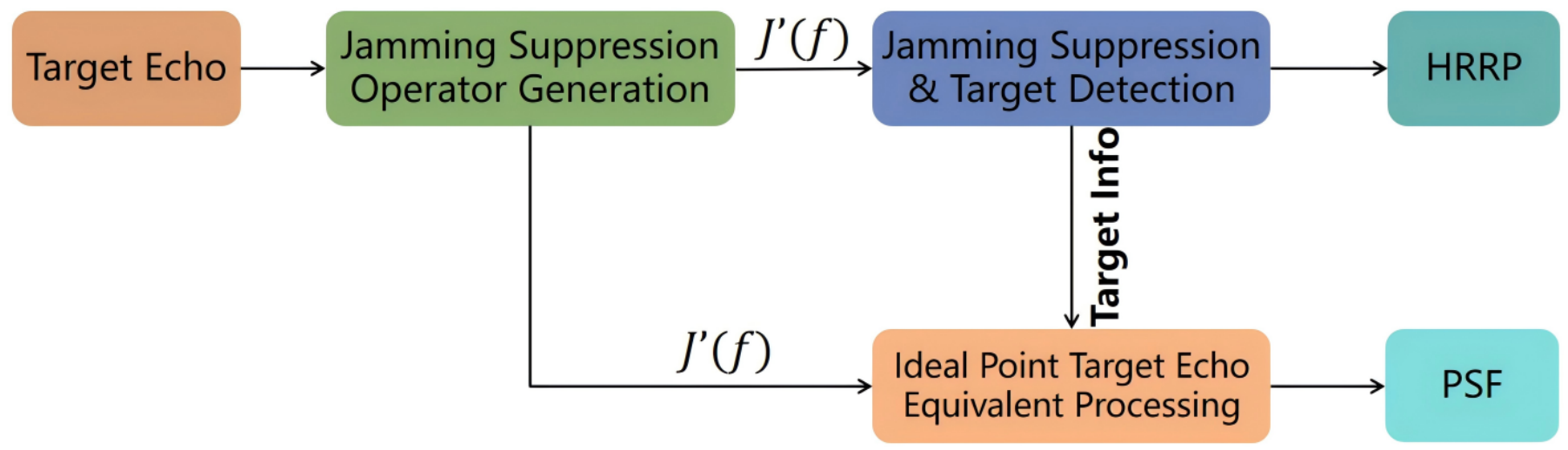}  
    \caption{Process of generating HRRP and prior jamming information.}
    \label{figure:PSF}
\end{figure}

\begin{figure}[bp]
    \centering
    \includegraphics[width=8.5cm,height=3cm]{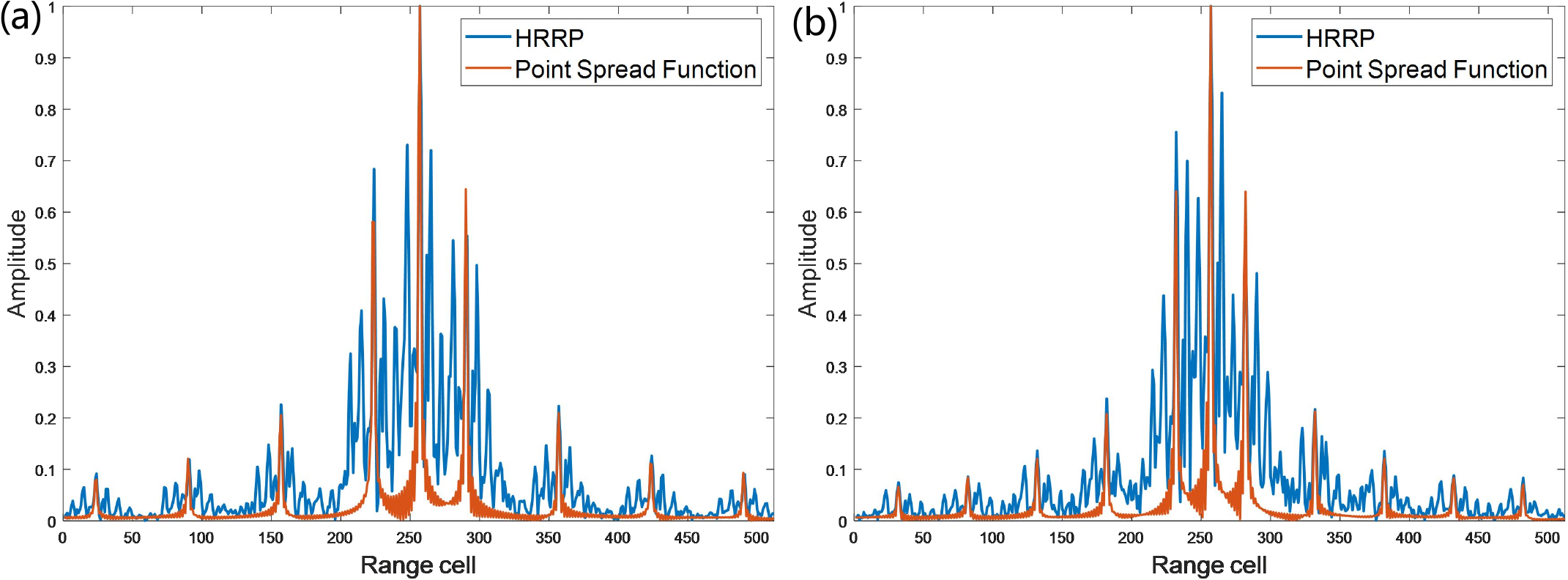}  
    \caption{Visualization of HRRP and the corresponding prior jamming information under ISRJ with different jamming configurations.}
    \label{figure:prior_plot}
\end{figure}

In radar imaging systems, the PSF characterizes the imaging performance of an ideal point target and is thus widely regarded as the impulse response of the radar sensing system~\cite{setlur2013multipath}. Motivated by this interpretation, we consider obtaining the PSF under ISRJ as a distortion indicator, which describes the distortion characteristics in the range dimension response of an ideal point target.
Specifically, we propose a fast prior information modeling method to construct the PSF under jamming. As illustrated in Figure~\ref{figure:PSF}, We first extract the jamming suppression operator \( J'(f) \) from the jammed target echo, where its frequency-domain period is \( T'_f \) and its frequency-domin pulse width is \( \tau'_f \), respectively. We then perform jamming suppression and target detection on the target echo to obtain target information (e.g., range and velocity), and subsequently generate the HRRP from the jamming-suppressed target echo. Based on the obtained target information, we generate the corresponding virtual ideal point target echo. This virtual echo is then processed using the operator \( J'(f) \) to perform equivalent jamming suppression, thereby yielding the PSF under the same jamming condition.

As shown in Figure~\ref{figure:prior_plot}, the distortion characteristics of the PSF closely resemble those observed in real target HRRP under ISRJ, effectively capturing the underlying degradation mechanism caused by jamming. By embedding this physically interpretable prior into the network, we enhance the model’s ability to distinguish between jamming artifacts and target-specific features.

\subsection{Feature Interaction Module}

To fully exploit prior information, we design a feature interaction module, as shown in Figure~\ref{fig:main}.
It comprises two components: a dual-attention feature fusion component for reciprocal HRRP-prior feature fusion, and a prior-guided feature selection component that extracts the most relevant HRRP features guided by prior information.

\subsubsection{\textbf{Dual-Attention Feature Fusion Module}}

In the first stage of the dual-attention feature fusion module, a self-attention mechanism~\cite{vaswani2017attention} is independently applied to both the HRRP features and the prior information features to capture their respective intra-feature dependencies and enhance their individual representational capabilities.
In the second stage, a bi-directional cross-attention mechanism is employed to facilitate deep mutual interaction between the two feature domains. As shown in Figure~\ref{fig:main}, we first perform HRRP-to-prior cross-attention, where HRRP features serve as the query and prior features as the key and value, enabling the model to selectively attend to relevant jamming-related information.
After completing the HRRP-to-prior cross-attention, a prior-to-HRRP cross-attention is performed in reverse, allowing the prior features to be refined based on the HRRP representation.
This dual-attention mechanism enables deep and reciprocal interactions between HRRP and prior features, thereby enhancing the model’s robustness and generalization under jamming conditions.

\subsubsection{\textbf{Prior-guided Feature Selection Module}}
To exploit prior jamming information for HRRP recognition, we propose a prior-guided feature selection module that learns intrinsic jamming patterns by selecting HRRP features most relevant to the prior.
As shown in Figure~\ref{fig:main}c, let \( F_{\text{HRRP}'} \in \mathbb{R}^{N_H \times d_k} \) and \( F_{\text{Prior}'} \in \mathbb{R}^{N_P \times d_k} \) denote HRRP and prior features, respectively. The objective is to select \( N_q \) representative HRRP feature tokens, guided by the prior information, to serve as the initial query inputs for the subsequent classification task. The \( \text{Top}_{N_q} \) indices \( \mathbf{I}_{N_q} \) are determined based on the cross-attention relevance between HRRP and prior features, computed as:

\begin{equation}
\mathbf{I}_{N_q} = \text{Top}_{N_q} \left( \text{Max}^{(-1)} \left( F_{\text{HRRP}'} F_{\text{Prior}'}^\top \right) \right),
\end{equation}
where the operator \( \text{Max}^{(-1)} \) performs a maximum operation along the last dimension. The \( \text{Top}_{N_q} \) denotes the operation of selecting the indices corresponding to the top \( N_q \) attention scores, and the selected indices \( \mathbf{I}_{N_q} \) are then used to select the most informative HRRP features. The selected features are then fed into the following classification network.

\subsection{Hybrid Loss Function}

\begin{figure}[tbp]
    \centering
    \includegraphics[width=8cm,height=3.1cm]{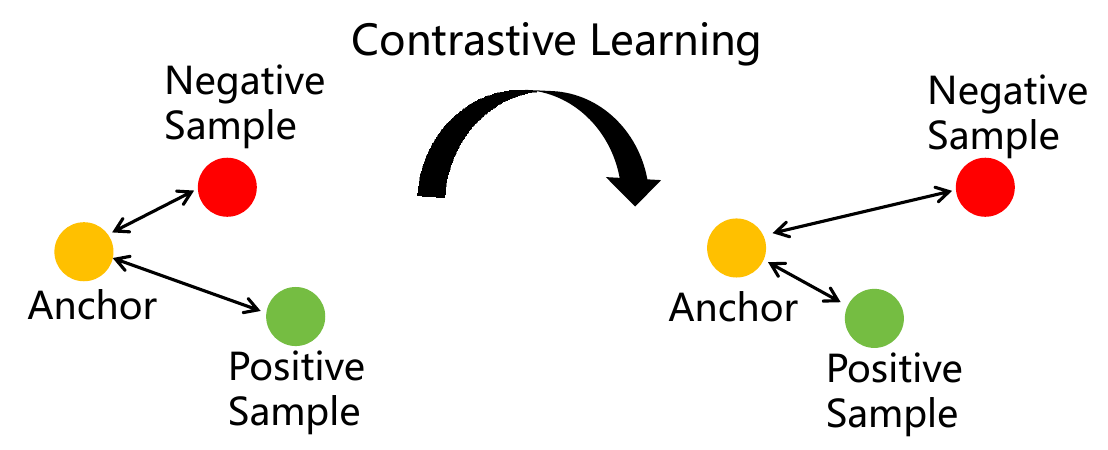}  
    \caption{Visualization of feature distributions learned through supervised contrastive learning.}
    \label{figure:SCL}
\end{figure}

To enhance feature discriminability and robustness under jamming, we employ a hybrid loss that combines cross-entropy and supervised contrastive loss during the training phase.
The cross-entropy loss is defined as:
\begin{equation}
\mathcal{L}_{\text{CEL}} = -\frac{1}{I} \sum_{i=1}^{I} \sum_{c=1}^{M} y_{ic} \log(p_{ic}),
\end{equation}
where \( I \) is the batch size, \( M \) is the number of classes, \( y_{ic} \) is the one-hot label, and \( p_{ic} \) is the predicted probability.
The supervised contrastive loss promotes intra-class compactness and inter-class separation~\cite{NEURIPS2020_d89a66c7}. Given a mini-batch of samples indexed by set \( I \), each sample \( i \in I \) is treated as an anchor. Samples with the same class label as \( i \) form its positive set \( P(i) \), while those with different labels form the negative set. The supervised contrastive loss is defined as:

\begin{small}
\begin{equation}
\mathcal{L}_{\text{SCL}} = \sum_{i \in I} \frac{1}{|P(i)|} \sum_{p \in P(i)} 
\log \frac{\exp(\text{sim}(z_i, z_p)/\tau)}{\sum_{a \in A(i)} \exp(\text{sim}(z_i, z_a)/\tau)},
\end{equation}
\end{small}
where \( z_i \) is the projected representation of sample \( i \), \( \text{sim}(\cdot) \) denotes cosine similarity, \( \tau \) is a temperature scaling factor, and \( A(i) \) includes all samples in the batch except \( i \). 
 
The final hybrid loss is computed as a weighted sum:
\begin{equation}
\mathcal{L}_{\text{total}} = \mathcal{L}_{\text{CEL}} + \lambda \cdot \mathcal{L}_{\text{SCL}},
\end{equation}
where \( \lambda \in [0, 1] \) balances the classification objective and representation learning.

\section{EXPERIMENTS}
\label{sec:experiments}

\subsection{Experimental Settings}
\label{ssec:Experimental Settings}

\subsubsection{\textbf{Dataset}}
To evaluate recognition performance, we construct both simulated and measured datasets. The simulated dataset contains six classes of aircraft, namely An-71, E-2D, B-2, F-16, F-22, and Tu-160. Each aircraft class contains 3,801 frequency-response samples obtained via the electromagnetic simulation software CST, resulting in a total of 22,806 samples. Based on these frequency-response samples, radar echoes are then simulated using an LFM waveform with bandwidth \(B=320~MHz\), pulse width \(T=100~\mu s\), and carrier frequency \(f_0=3~GHz\). By superimposing ISRJ signals with sampling periods \(\tau_s \in [0.5, 3.2]~\mu\text{s}\) at an interval of \(\Delta\tau_s=0.3~\mu\text{s}\) and retransmission periods \(\tau_t \in \{\tau_s, 2\tau_s\}\) onto radar echoes, an HRRP dataset is generated under 21 distinct jamming parameter configurations (including one configuration without jamming), with each configuration containing 22,806 HRRP samples.
 The measured dataset consists of five types of civil aircraft: Airbus A320 (1,527 samples), Airbus A319 (1,651 samples), Boeing 777 (730 samples), Boeing 737 (1,044 samples), and Bombardier CRJ900 (709 samples), resulting in 5,661 HRRP samples per jamming configuration under the same radar and jamming settings as the simulated dataset.

For both datasets, HRRP samples under 15 jamming parameter configurations are selected to evaluate the in-distribution (InD) performance, and the samples are randomly divided into training and testing subsets with a ratio of 8:2 according to observation angles. Meanwhile, HRRP samples under the remaining six unseen jamming parameter configurations are utilized to construct an independent out-of-distribution (OOD) test set. This OOD set is employed to evaluate the model’s generalization ability under unseen jamming parameters.

\subsubsection{\textbf{Hyperparameter Details}}
In the Encoder, a four-layer 1D CNN architecture with channel-spatial attention~\cite{Woo_2018_ECCV} is adopted. The dual-attention feature fusion module adopts a one-layer self-attention and a two-layer cross-attention mechanism, each with embedding dimension $d_k=64$, FFN hidden size of 128, and 8 attention heads. In the prior-guided feature selection module, the feature dimensions are set to $N_H = N_P = 64$, with $N_q = 48$ query tokens. For the hybrid loss function, the weight and temperature parameters are set to $\lambda = 0.4$ and $\tau = 0.1$, respectively. Optimization is performed using Adam with a learning rate of 0.001, a batch size of 512, and 50 training epochs. All experiments are conducted on an NVIDIA GeForce RTX 4090 GPU.

\subsection{Recognition Performance}

To evaluate the recognition performance of the proposed method, several representative models including TARAN~\cite{xu2019target}, TACNN~\cite{chen2022target}, SAN~\cite{10661229}, ConvLSTM~\cite{9293131}, and ITAViT~\cite{gao2024interpretable}, are reproduced as baseline comparison methods. In addition, the Accuracy, Precision, Recall, and F1-score are utilized as quantitative metrics to quantify the recognition performance.

\begin{table}[tbp]
\centering
\caption{Recognition performance comparison under ISRJ (InD \& OOD).}

{
\renewcommand{\arraystretch}{1} 
\setlength{\tabcolsep}{3pt} 
\begin{tabular}{lcccccccc}
\toprule
\multirow{2}{*}{\textbf{Model}} &
\multicolumn{4}{c}{\textbf{InD (\%)}} & 
\multicolumn{4}{c}{\textbf{OOD (\%)}} \\
\cmidrule(lr){2-5} \cmidrule(lr){6-9}
& Acc. & Prec. & Rec. & F1. & Acc. & Prec. & Rec. & F1. \\
\midrule

\multicolumn{9}{c}{\textit{\textbf{Simulated Dataset}}} \\
TARAN      & 87.73 & 88.06 & 87.73 & 87.79 & 64.71 & 64.57 & 64.71 & 64.55 \\
TACNN      & 93.27 & 93.39 & 93.27 & 93.28 & 65.96 & 66.75 & 65.96 & 65.87 \\
SAN        & 93.43 & 93.46 & 93.43 & 93.44 & 70.51 & 71.42 & 70.51 & 70.14 \\
ConvLSTM   & 90.49 & 90.65 & 90.49 & 90.53 & 67.72 & 67.97 & 67.72 & 66.86 \\
ITAViT     & 94.61 & 94.62 & 94.60 & 94.61 & 74.76 & 77.41 & 74.76 & 74.57 \\
\textbf{Ours} & \textbf{97.17} & \textbf{97.19} & \textbf{97.17} & \textbf{97.17} & \textbf{84.28} & \textbf{84.60} & \textbf{84.28} & \textbf{84.26} \\
\midrule
\multicolumn{9}{c}{\textit{\textbf{Measured Dataset}}} \\
TARAN      & 92.57 & 93.34 & 93.05 & 93.18 & 73.63 & 76.88 & 72.01 & 73.71 \\
TACNN      & 94.54 & 95.25 & 95.05 & 95.14 & 78.86 & 83.90 & 76.98 & 79.48 \\
SAN        & 95.04 & 95.44 & 95.65 & 95.54 & 81.78 & 83.99 & 80.79 & 82.05 \\
ConvLSTM   & 88.33 & 89.02 & 89.53 & 89.24 & 71.95 & 73.05 & 72.29 & 72.06 \\
ITAViT     & 94.22 & 94.91 & 94.75 & 94.81 & 80.31 & 83.40 & 79.11 & 80.79  \\
\textbf{Ours} & \textbf{97.26} & \textbf{97.27} & \textbf{97.71} & \textbf{97.49} & \textbf{89.00} & \textbf{90.74} & \textbf{88.33} & \textbf{89.40} \\
\bottomrule
\end{tabular}
}

\label{tab:recognition_combined_ood}
\end{table}

\begin{figure}[bp]
    \centering
    \includegraphics[width=0.47\textwidth]{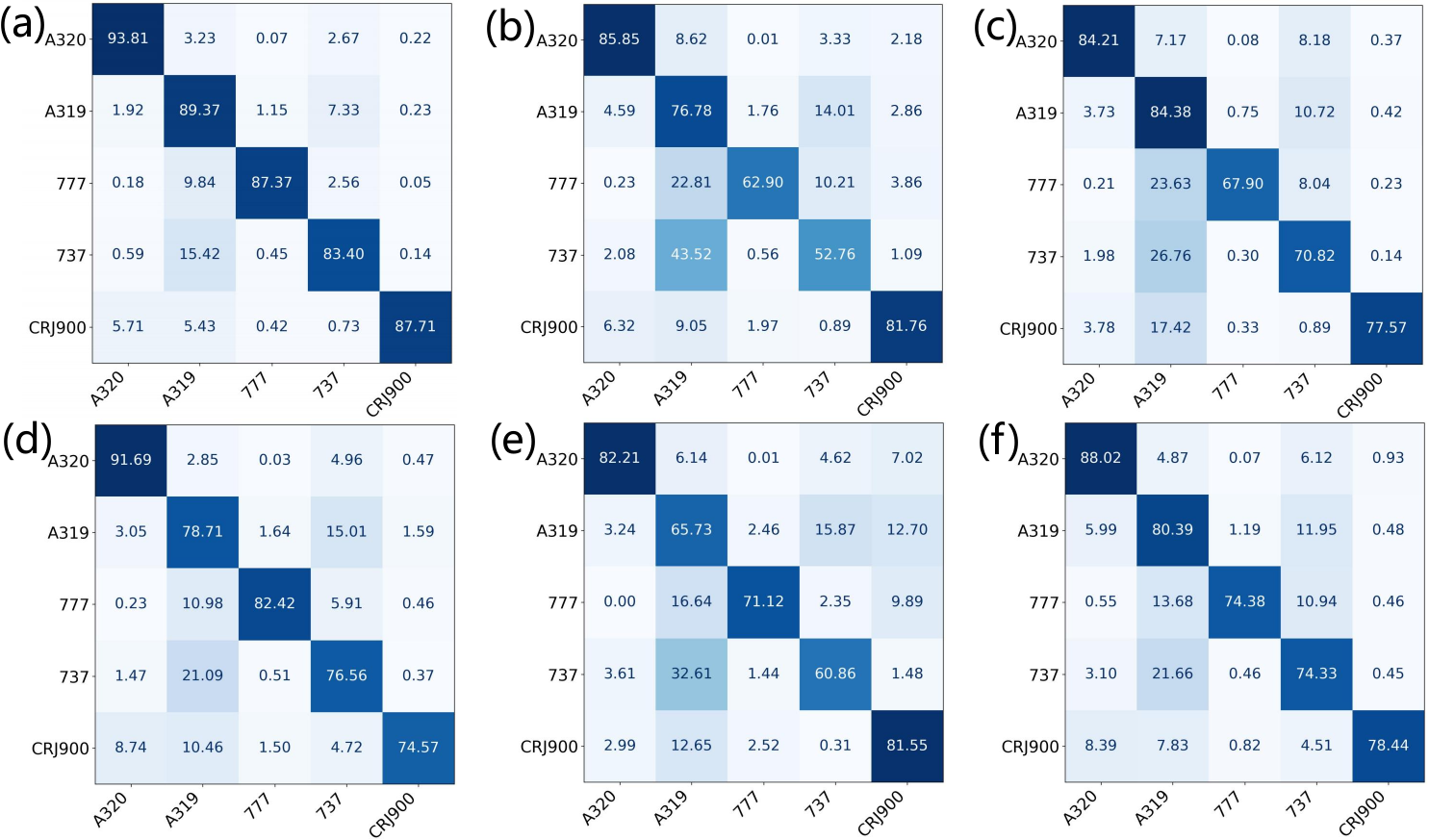}
    
    \caption{Confusion matrices of the measured dataset (OOD). (a) Ours. (b) TARAN. (c) TACNN. (d) SAN. (e) ConvLSTM. (f) ITAViT.}
    \label{fig:confusion_matrix_real}
\end{figure}

As shown in Table \ref{tab:recognition_combined_ood}, the proposed method achieves the highest recognition performance on both simulated and measured datasets. Specifically, under the InD condition, our method attains an Accuracy of 97.17\% on the simulated dataset and 97.26\% on the measured dataset, outperforming all comparison methods. More importantly, our method exhibits outstanding generalization ability under the OOD condition, achieving an Accuracy of 84.28\% on the simulated dataset and 89.00\% on the measured dataset. These results surpass the second-best model by 9.52\% and 8.69\%, respectively.

\begin{figure}[tbp]
    \centering
    \includegraphics[width=0.47\textwidth]{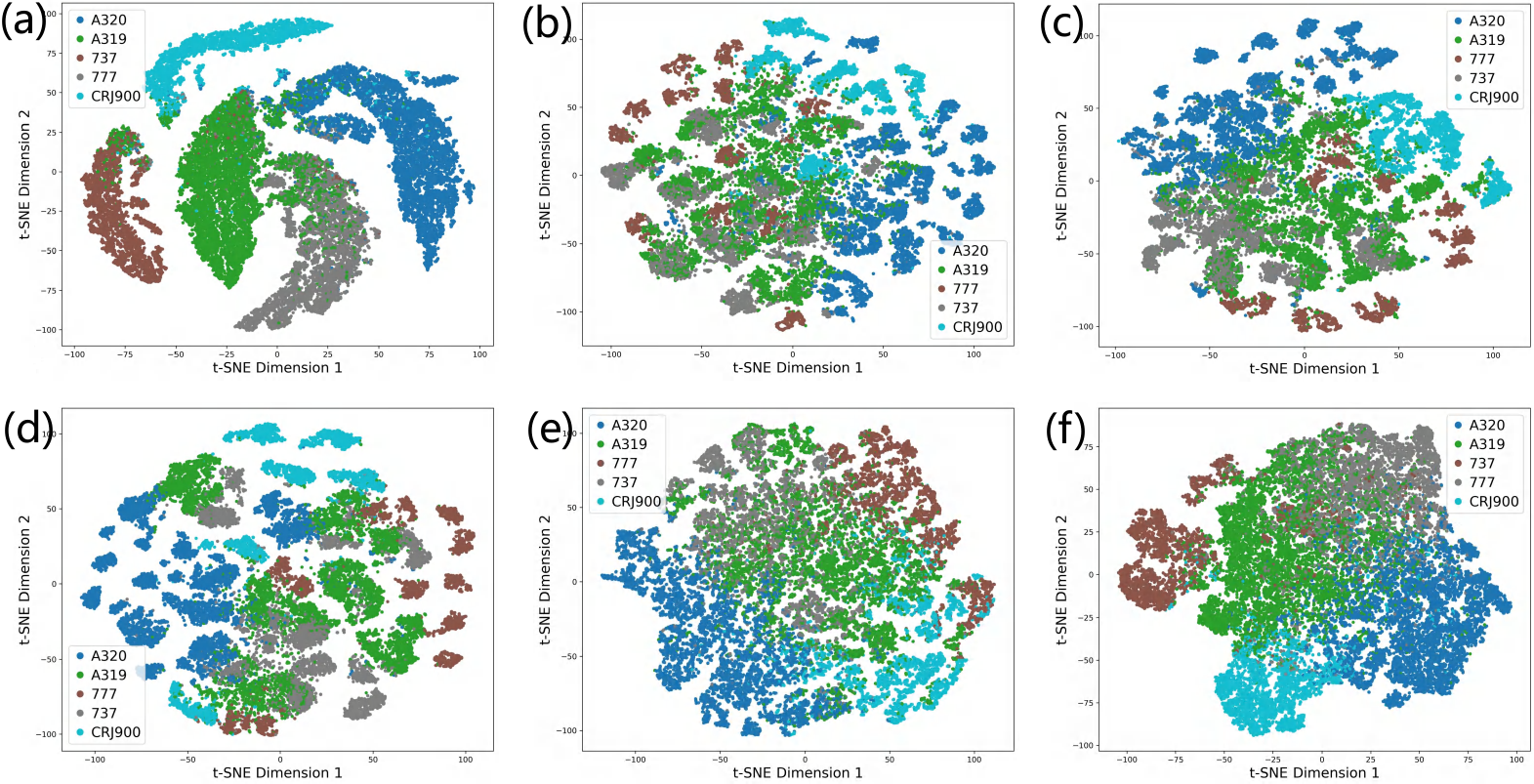}
    
    \caption{t-SNE visualizations of the measured dataset (OOD). (a) Ours. (b) TARAN. (c) TACNN. (d) SAN. (e) ConvLSTM. (f) ITAViT.}
    \label{fig:tsne_real}
\end{figure}

Figure~\ref{fig:confusion_matrix_real} presents the confusion matrices of the measured dataset (OOD). It shows that our method exhibits the least confusion among all models. Notably, our method achieves an Accuracy exceeding 89\% for both A320 and A319. Moreover, it consistently maintains an Accuracy above 83\% across all classes, which is substantially higher than that of other methods. 
Figure~\ref{fig:tsne_real} presents the t-SNE visualizations of the measured dataset (OOD) to display the feature extraction performance. It can be observed that features from our method form compact and well-separated clusters, whereas those from baseline methods are more dispersed with noticeable class overlap.

These results highlight the superior generalization ability of the proposed method. With the aid of prior information, the model can learn invariant features from distorted HRRP and maintain robust recognition performance across different jamming parameters, thus better meeting the requirements of practical applications.

\subsection{Ablation Study}
\begin{table}[bp]
\centering
\caption{Ablation study results (InD \& OOD).}
\label{tab:ablation_comparison_ood}
{
\setlength{\tabcolsep}{3pt} 
\renewcommand{\arraystretch}{1} 
\begin{tabular}{lcccccccc}
\toprule
\multirow{2}{*}{\textbf{Model}} &
\multicolumn{4}{c}{\textbf{InD (\%)}} & 
\multicolumn{4}{c}{\textbf{OOD (\%)}} \\
\cmidrule(lr){2-5} \cmidrule(lr){6-9}
& Acc. & Prec. & Rec. & F1. & Acc. & Prec. & Rec. & F1. \\
\midrule

\multicolumn{9}{c}{\textit{\textbf{Simulated Dataset}}} \\
Full Model & \textbf{97.17} & \textbf{97.19} & \textbf{97.17} & \textbf{97.17} & \textbf{84.28} & \textbf{84.60} & \textbf{84.28} & \textbf{84.26} \\
w/o Prior             & 96.99 & 97.01 & 97.00 & 96.99 & 80.58 & 81.12 & 80.58 & 80.40 \\
w/o HLF             & 96.81 & 96.83 & 96.81 & 96.81 & 79.65 & 79.98 & 79.65 & 79.55 \\
w/o Both           & 96.39 & 96.42 & 96.39 & 96.39 & 76.40 & 76.88 & 76.40 & 76.29 \\
\midrule

\multicolumn{9}{c}{\textit{\textbf{Measured Dataset}}} \\
Full Model & \textbf{97.26} & \textbf{97.27} & \textbf{97.71} & \textbf{97.49} & \textbf{89.00} & \textbf{90.74} & \textbf{88.33} & \textbf{89.40} \\
w/o Prior             & 96.68 & 96.60 & 97.21 & 96.89 & 85.33 & 86.75 & 84.73 & 85.48 \\
w/o HLF             & 96.81 & 97.08 & 97.33 & 97.19 & 85.86 & 87.09 & 85.01 & 85.87 \\
w/o Both           & 96.40 & 96.59 & 97.05 & 96.81 & 81.93 & 83.89 & 81.64 & 82.32 \\
\bottomrule
\end{tabular}
}
\end{table}

To further validate the effectiveness of each module in the proposed algorithm, we conduct ablation studies on two key components: the prior information (Prior) and the hybrid loss function (HLF). Specifically, in w/o Prior, the prior features are substituted with the HRRP features in the feature interaction module to assess the contribution of prior information. In w/o HLF, the weight coefficient \(\lambda\) in the hybrid loss function is set to 0, leaving only the cross-entropy loss to examine the effectiveness of the supervised contrastive loss. In w/o Both, the prior features are substituted with HRRP features in the feature interaction module, while the weight coefficient \(\lambda\) in the hybrid loss function is set to 0.

\begin{figure}[tbp]
    \centering
    \includegraphics[width=0.47\textwidth]{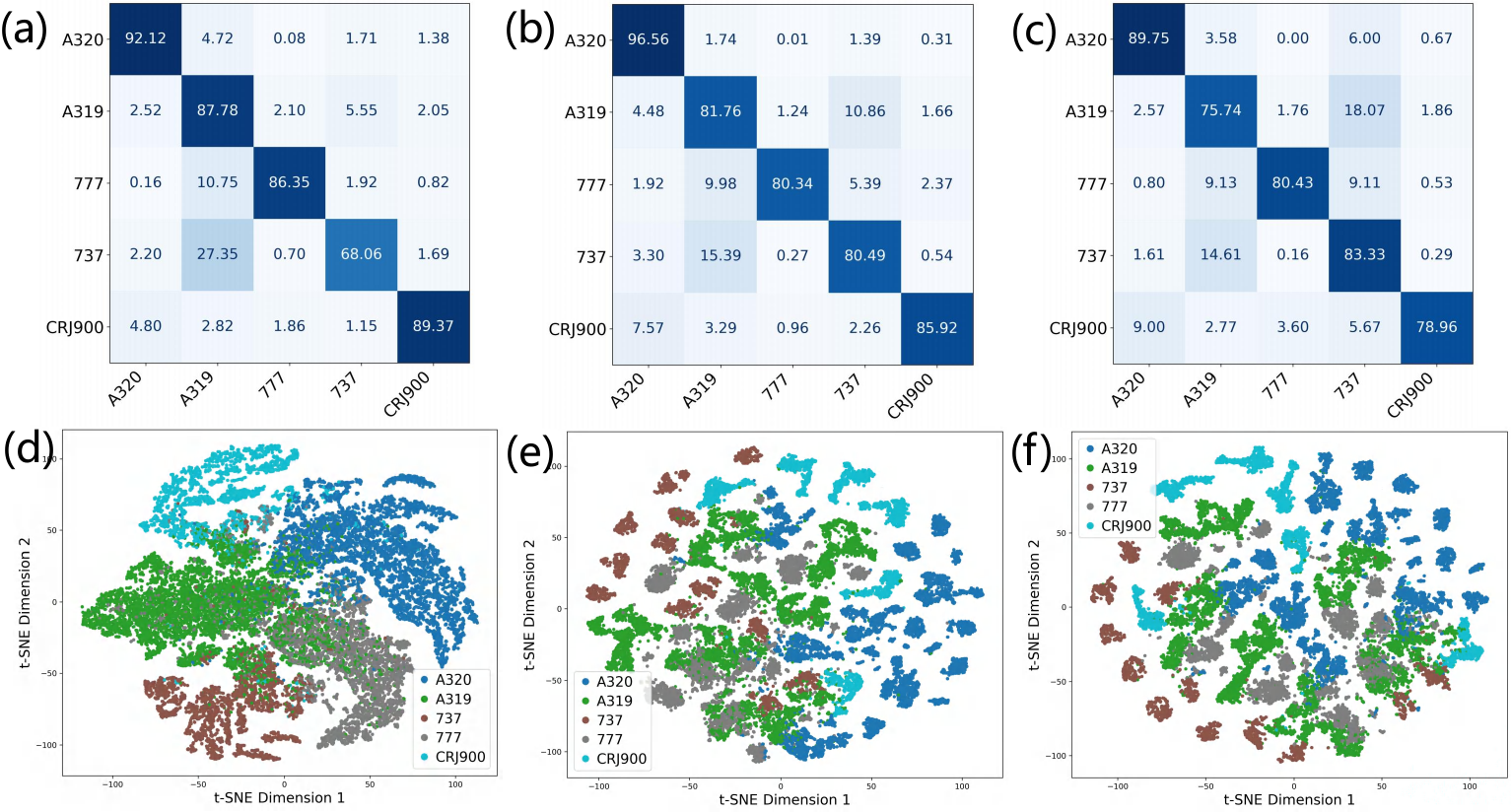}
    
    \caption{Visualizations of the ablation study on the measured dataset (OOD). (a)-(c) confusion matrices; (d)-(f) t-SNE visualizations. (a,d) w/o Prior; (b,e) w/o HLF; (c,f) w/o Both.}
    \label{fig:ablation_real}
\end{figure}

As shown in Table \ref{tab:ablation_comparison_ood}, both the prior information and the hybrid loss function notably enhance model performance, especially under the OOD condition. Removing either module leads to a clear drop in recognition performance, while removing both causes the largest degradation. 
As illustrated in Figure~\ref{fig:ablation_real}, the confusion matrices indicate higher inter-class confusion in these cases. Correspondingly, the t-SNE visualizations show that removing the prior information leads to more dispersed feature distributions and increased class overlap, while removing the hybrid loss function causes the features under different jamming parameters to fragment into discrete small clusters. These results confirm that prior information improves generalization capability of the proposed method, and the hybrid loss function strengthens feature discriminability under distribution shifts.

\section{Conclusion}
\label{sec:conclusion}
In this paper, we propose a novel HRRP recognition network under ISRJ by incorporating prior jamming information, which is modeled via a virtual PSF, into a deep neural network. To the best of our knowledge, this is the first study that models jamming-induced distortions in HRRP from a PSF perspective, effectively bridging physical signal modeling and data-driven learning. The proposed method introduces three key components: a prior jamming information modeling method, a prior-guided feature interaction module, and a hybrid loss function, collectively enabling robust and highly generalizable HRRP recognition under ISRJ conditions. Notably, our method establishes a general framework for robust target recognition under jamming conditions, laying a foundation for its extension to broader jamming types and real-world deployments.
\bibliographystyle{ieeetr}
\bibliography{refs}
\end{document}